\pdfoutput=1

\RequirePackage{fix-cm}

\documentclass[onecolumn,numbers,sort&compress]{svjour3}

\smartqed

\usepackage{graphicx}
\usepackage{amssymb,amsmath}
\usepackage{txfonts}
\usepackage{natbib}
\usepackage{hyperref}

\hypersetup{pdftitle = The title of my PDF, pdfauthor = My name, pdfsubject= The subject, pdfkeywords = keyword1 keyword2 keyword3}
\hypersetup{colorlinks = true, linkcolor = blue, anchorcolor = red, citecolor = blue, filecolor = red, pagecolor = red, urlcolor = blue}

\journalname{International Journal of Applied and Computational Mathematics}

\newcommand{\mathz}{\ooalign{$z$\cr\hfil\rule[.5ex]{.2em}{.06ex}\hfil\cr}}

\begin{document}

\title{On the convergence dynamics of the Sitnikov problem with non-spherical primaries}

\author{Euaggelos E. Zotos \and Md Sanam Suraj \and Rajiv Aggarwal \and Amit Mittal}

\institute{Euaggelos E. Zotos: \at
             Department of Physics, School of Science, \\
             Aristotle University of Thessaloniki, \\
             GR-541 24, Thessaloniki, \\
             Greece \\
             \email{\url{evzotos@physics.auth.gr}}
         \and
           Md Sanam Suraj: \at
              Department of Mathematics, Sri Aurobindo College, \\
              University of Delhi, New Delhi-110017 \\
              India \\
             \email{\url{mdsanamsuraj@gmail.com}}
         \and
           Rajiv Aggarwal: \at
             Department of Mathematics, Deshbandhu College, \\
             University of Delhi, New Delhi-110019 \\
             India \\
             \email{\url{rajiv_agg1973@yahoo.com}}
         \and
           Amit Mittal: \at
             Department of Mathematics, ARSD College, \\
             University of Delhi, New Delhi-110021 \\
             India \\
             \email{\url{to.amitmittal@gmail.com}}
}

\date{Published online: 14 March 2019}

\titlerunning{Convergence properties of the Sitnikov problem with non-spherical primaries}

\authorrunning{E.E. Zotos et al.}

\maketitle

\begin{abstract}
We investigate, using numerical methods, the convergence dynamics of the Sitnikov problem with non-spherical primaries, by applying the Newton-Raphson (NR) iterative scheme. In particular, we examine how the oblateness parameter $A$ influences several aspects of the method, such as its speed and efficiency. Color-coded diagrams are used for revealing the convergence basins on the plane of complex numbers. Moreover, we compute the degree of fractality of the convergence basins on the complex space, as a relation of the oblateness, by using different computational tools, such the fractal dimension as well as the (boundary) basin entropy.


\end{abstract}

\defcitealias{ZSAS18}{Paper I}

\section{Introduction}
\label{intro}

The simplest scenario, according to which the two primary bodies perform planar circular orbits (with zero eccentricity, $e = 0$), is known as the MacMillan problem \cite{M11}, while this notion was originally introduced in \cite{P07}. Moreover, the first qualitative results on the Sitnikov problem have been conducted in \cite{S60} and \cite{M73}.

Knowing the exact coordinates of the points of equilibrium of a system is an important issue. However, this is not possible for many complicated dynamical systems for which there are no analytical equations for the positions of the equilibrium points. This automatically means that only by using numerical methods we can obtain the locations of the libration points. As we know, in all numerical methods the initial conditions are very important. Indeed, for some starting points the numerical methods may converge relatively fast to a root, while for other initial conditions they may require a considerable amount of iterations. Usually, points with fast convergence belong to the basins of convergence (Boc), while slow converging points are situated in the vicinity of the fractal basin boundaries. Therefore, it is of great importance to know the location of the Boc for a dynamical system, because then we automatically are aware of the optimal initial conditions for the numerical methods. Here, we would like to point out that the Boc of a dynamical system strongly depend on the chosen numerical method. In other words, different numerical methods yield to completely different Boc, for the same dynamical system (see e.g., \cite{Z17,Z18}).

The Boc define complicated geometrical structures on the complex or the configuration $(x,y)$ plane. Another important aspect is knowing the degree of fractality of the convergence regions. A quantitatively estimation of the degree of fractality can be easily achieved by computing several numerical indicators, such as the uncertainty or fractal dimension (see e.g., \cite{AVS01,AVS09}) or the basin entropy (see e.g., \cite{DWGGS16}). Both these quantities can provide safe results, regarding the degree of fractality of a dynamical system.

Our article has the following layout: in Section \ref{sys} we describe the mathematical formulation of the dynamical model. The next Section \ref{numres} contains all the numerical outcomes of our work, about the properties of the  Sitnikov problem with non-spherical primaries. In the last Section \ref{conc}, we emphasize the conclusions of our computational analysis.

\section{Mathematical formulation of the system}
\label{sys}

The system consists of two primaries whose dimensionless masses are $m_1 = \mu$ and $m_2 = 1 - \mu$, where $\mu = m_2/(m_1 + m_2) \leq 1/2$ is the well known mass parameter \citep{S67}. The centres of the two primaries lie on the horizontal $Ox$ axis and in particular at $(x_1, 0, 0)$ and $(x_2, 0, 0)$, where $x_1 = - \mu$ and $x_2 = 1 - \mu$. For each primary it is assumed that its shape resembles a spheroid, according to the value of the corresponding oblateness $A_i$, $i = 1,2$.

According to \cite{AS06,DM06,SSR75}, the function of the potential of the restricted circular problem with two oblate primaries is given by
\begin{equation}
\Omega(x,y,z) = \sum_{i=1}^{2} \frac{m_i}{r_i}\left(1 + \frac{A_i}{2r_i^2} - \frac{3A_i z^2}{2r_i^4}\right) + \frac{n^2}{2} \left(x^2 + y^2 \right),
\label{pot}
\end{equation}
where
\begin{align}
r_1 &= \sqrt{\left(x - x_1 \right)^2 + y^2 + z^2}, \nonumber\\
r_2 &= \sqrt{\left(x - x_2 \right)^2 + y^2 + z^2},
\label{dist}
\end{align}
are the respective distances between the test particle (third body), and the centers of the two primaries. Moreover, the mean motion is
\begin{equation}
n = \sqrt{1 + 3\left(A_1 + A_2 \right)/2}.
\label{mn}
\end{equation}

The third body of the system moves according to the following equations
\begin{align}
\ddot{x} &- 2 n \dot{y} = \frac{\partial \Omega}{\partial x}, \nonumber\\
\ddot{y} &+ 2 n \dot{x} = \frac{\partial \Omega}{\partial y}, \nonumber\\
\ddot{z} &= \frac{\partial \Omega}{\partial z}.
\label{eqmot}
\end{align}

For this system there is only one known motion integral, which reads
\begin{equation}
J = 2\Omega(x,y,z) - \left(\dot{x}^2 + \dot{y}^2 + \dot{z}^2 \right) = C.
\label{ham}
\end{equation}

\begin{figure}[!t]
\centering
\resizebox{0.6\hsize}{!}{\includegraphics{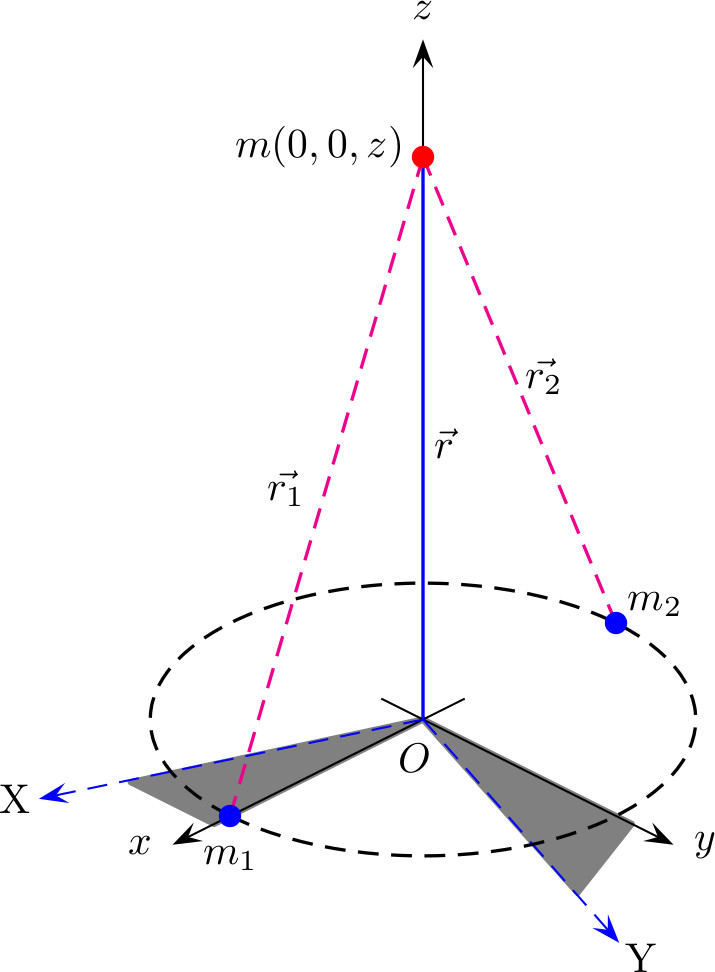}}
\caption{The space configuration of the Sitnikov problem, where two primary bodies with equal masses $(m_1 = m_2 = 1/2)$ perform circular orbits.}
\label{conf}
\end{figure}

By setting $x = y = 0$, $\mu = 1/2$, and $A_1 = A_2 = A$ in Eq. (\ref{pot}) we obtain the potential function of the Sitnikov problem
\begin{equation}
\Omega(z) = \frac{1}{r} + \frac{A}{2r^3} - \frac{3Az^2}{2r^5},
\label{potz}
\end{equation}
where $r = \sqrt{z^2 + 1/4}$.

Thus, the vertical motion, along the $z$ axis, of the test particle is governed by the following equation
\begin{equation}
\ddot{z} = - \frac{z}{r^3} - \frac{9Az}{2r^5} + \frac{15Az^3}{2r^7},
\label{eqmotz}
\end{equation}
while we can deduce that the corresponding Jacobi integral becomes
\begin{equation}
J(z,\dot{z}) = 2 \Omega(z) - \dot{z}^2 = C_{z}.
\label{hamz}
\end{equation}

In \citetalias{ZSAS18} we demonstrated that the points of equilibrium (roots) of the Sitnikov problem can be obtained through the equation of mition (\ref{eqmotz}). In addition, we seen that the value of the oblateness $A$ greatly influences the nature of the roots. Specifically:
\begin{itemize}
  \item For $A < -1/18$ there exist two pairs of real and imaginary roots.
  \item For $A = -1/18$ there exists a pair of imaginary roots.
  \item For $A \in (-1/18,0)$ there exist two pairs of imaginary roots.
  \item For $A = 0$ the root $\mathz = 0$ is the only root.
  \item For $A \in (0, 5/6)$ there exist two pairs of complex roots.
  \item For $A = 5/6$ there exists a pair of real roots.
  \item For $A > 5/6$ there exist two pairs of real roots.
\end{itemize}
We also concluded that the levels $A = \{-1/18, 0, 5/6 \}$ are in fact critical levels of the oblateness.

\section{Numerical results of the basins of convergence}
\label{numres}

The Boc on the plane of complex numbers can be determined by means of the numerical method of Newton-Raphson (NR). In \citetalias{ZSAS18} we shown that the corresponding iterative scheme reads

\begin{equation}
\mathz_{n+1} = \frac{12\mathz^3\left(A \left(50 - 80\mathz^2\right) + \left(1 + 4\mathz^2\right)^2 \right)}{128\mathz^6 + 48\mathz^4 - 6A \left(128\mathz^4 - 96\mathz^2 + 3 \right) - 1}.
\label{nr}
\end{equation}
At this point, we should emphasize that from now on the coordinate $z$ is treated as a complex variable $\mathz$, while the same approach was also successfully followed in \cite{DKMP12,Z17,Z18,ZSAS18}. In \cite{D10} it was demonstrated that the use of complex variables is necessary, because all the beautiful and impressive Boc, with the basin boundaries with fractal-like geometry, appear only on the plane of complex numbers.

\begin{figure*}[!t]
\centering
\resizebox{\hsize}{!}{\includegraphics{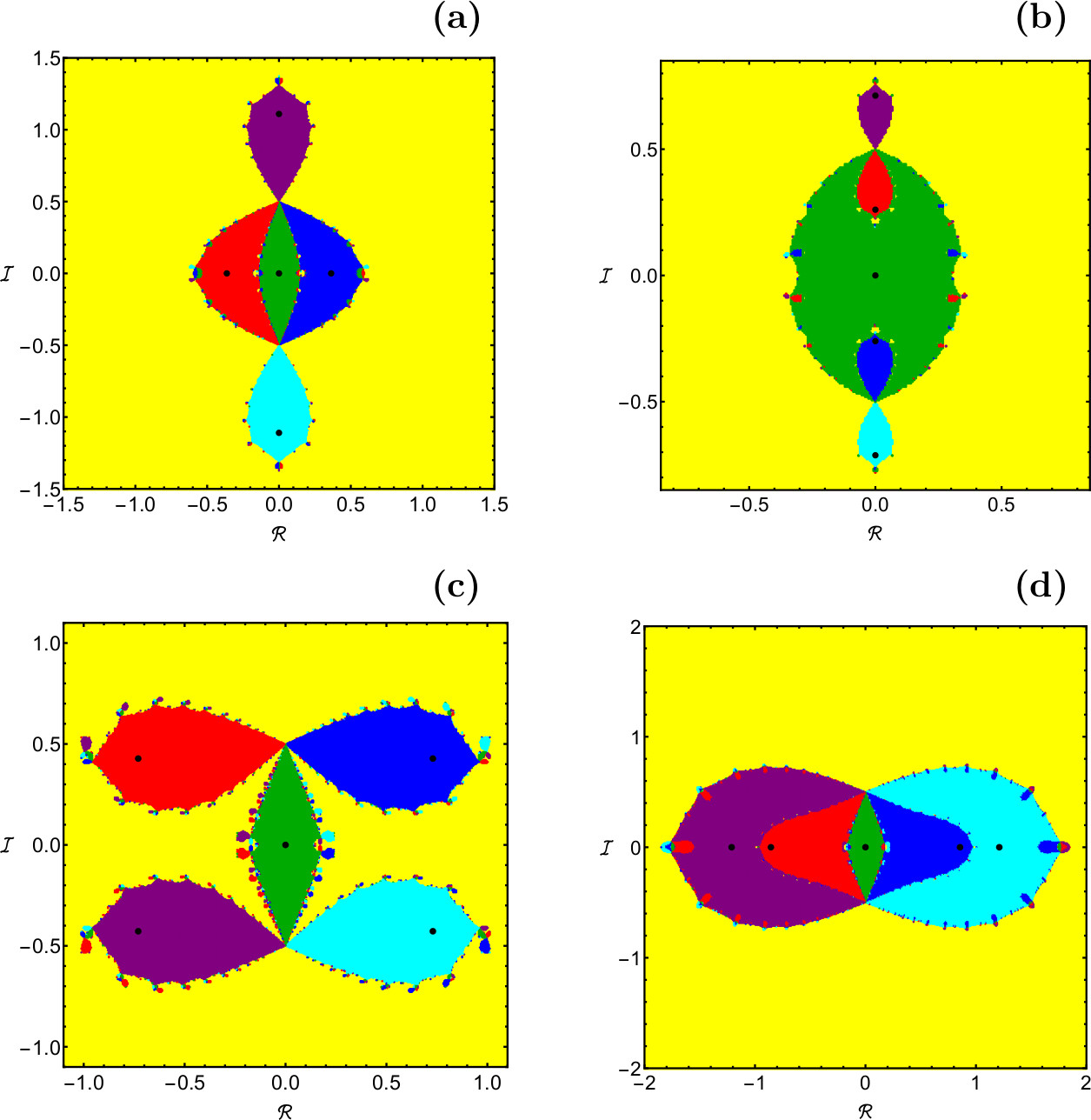}}
\caption{The NR Boc on the plane of complex numbers, for four characteristic values of the oblateness $A$. (a-upper left): $A = -0.2$; (b-upper right): $A = -0.025$; (c-lower left): $A = 0.4$; (d-lower right): $A = 0.9$. The five colors (red, blue, green, cyan, purple) indicate the five numerical attractors (roots) of the system. With black dots we denote, in each case, the positions of the roots. (Color figure online).}
\label{bas}
\end{figure*}

\begin{figure*}[!t]
\centering
\resizebox{\hsize}{!}{\includegraphics{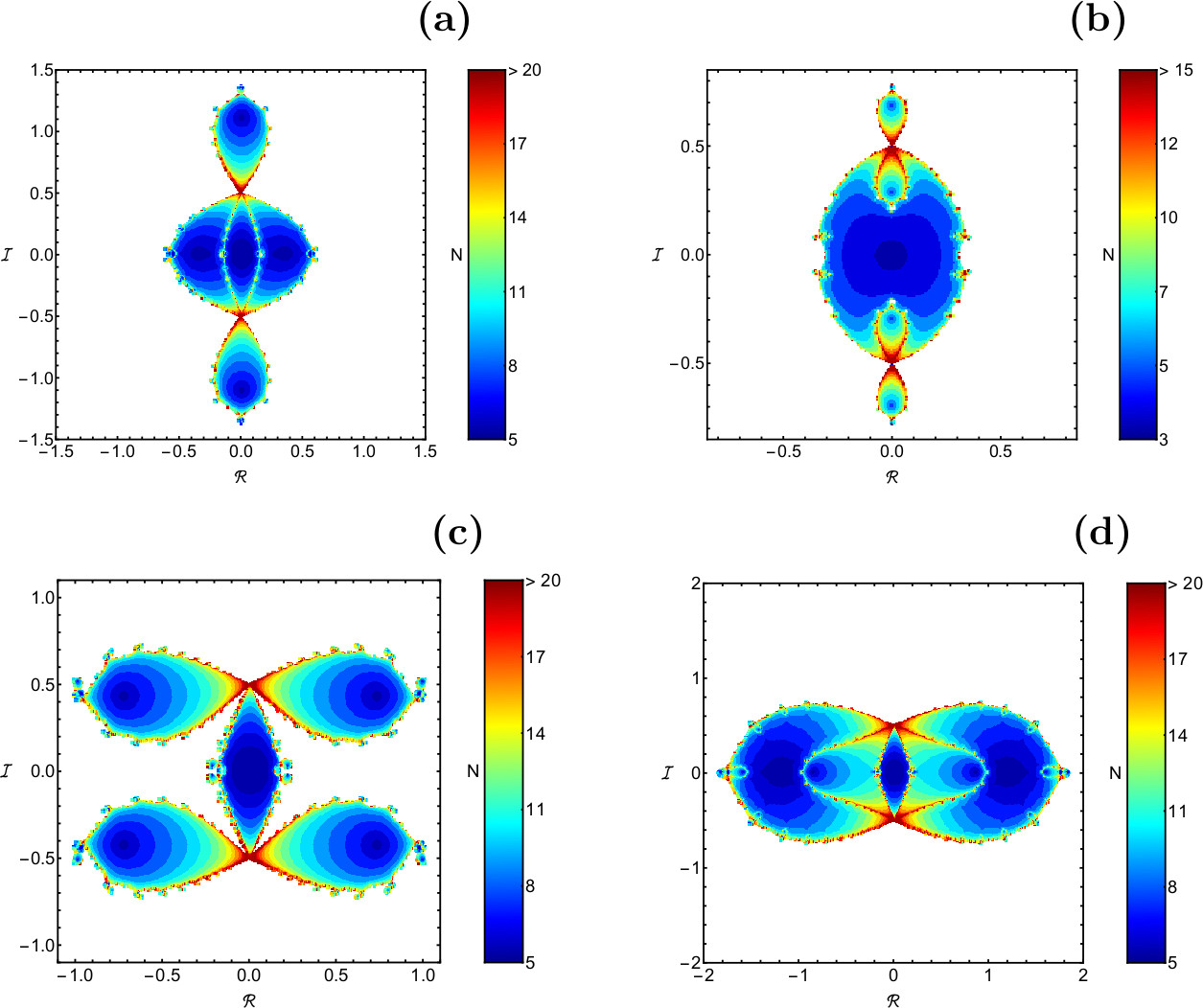}}
\caption{The distributions of the required iterations $N$, shown in parts (a-d) of Fig. \ref{bas}. (Color figure online).}
\label{iters}
\end{figure*}

\begin{figure*}[!t]
\centering
\resizebox{\hsize}{!}{\includegraphics{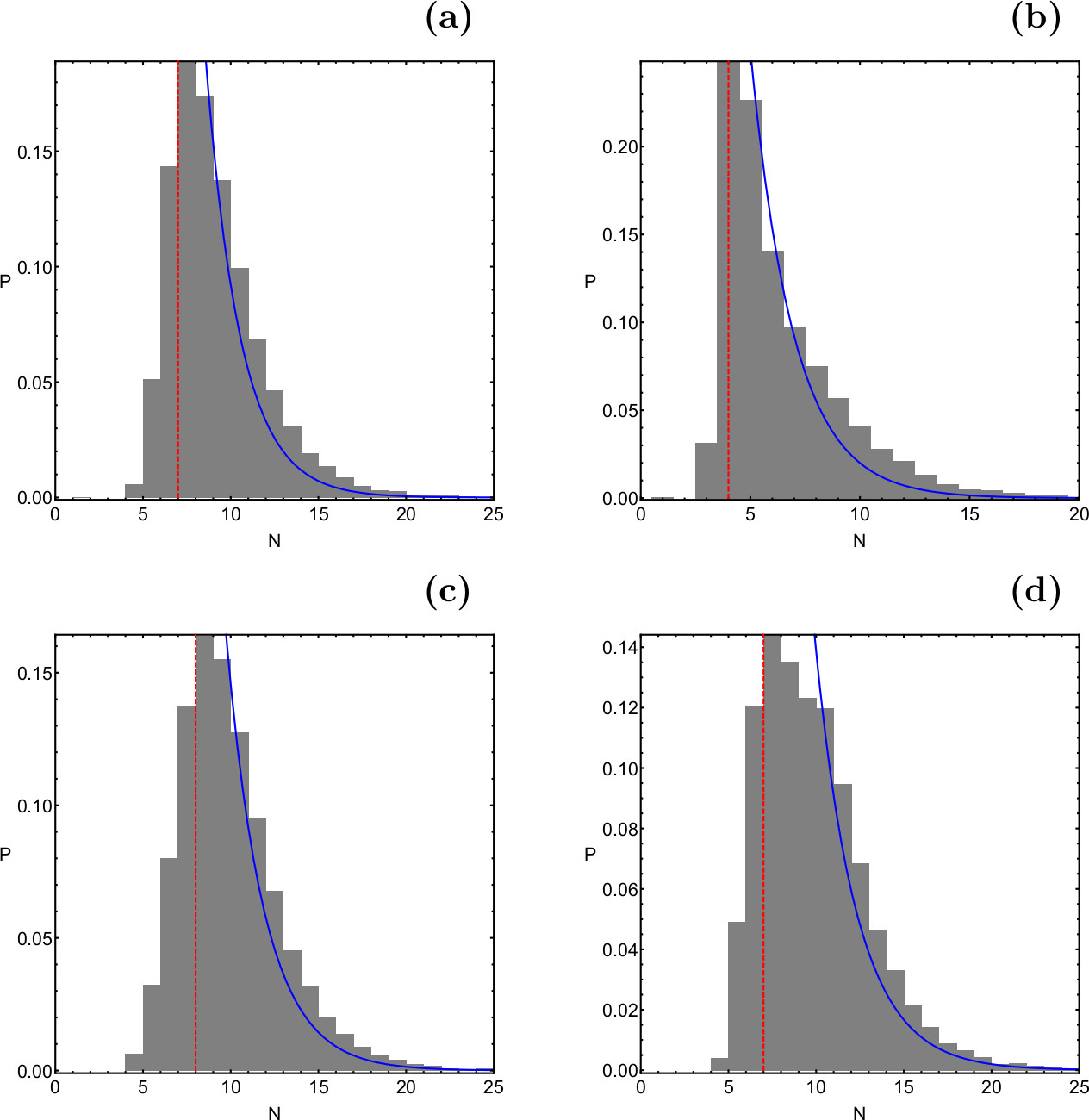}}
\caption{Histograms showing the distributions of the probability, for the four cases of Fig. \ref{bas}. With red, dashed, vertical lines we indicate the most probable number of iterations $N^{*}$, while with blue, solid lines we denote the best fit (Laplace probability density function) of the histograms. (Color figure online).}
\label{hist}
\end{figure*}

In \citetalias{ZSAS18} we presented in detail the structure of the Boc in the four intervals of $A$. In Fig. \ref{bas} we remind to the reader the structure as well as the geometry and the shape of the convergence regions, for four characteristic values of the oblateness $A$. It is seen, that in all cases the Boc have finite area. Moreover, the majority of the plane of complex numbers is occupied by initial conditions (yellow regions) for which the NR iterative method quickly diverges to very large complex numbers, thus numerically indicating divergence to infinity. The distributions of the required iterations $N$ are presented in Fig. \ref{iters}. It is evident that for initial conditions $(\mathcal{R}, \mathcal{I})$ near the roots the required iterations are low $(N \approx 5)$, while near the boundaries of the basins the NR iterative scheme needs more than 15 iterations for obtaining a root, with the predefined accuracy.

Fig. \ref{hist}(a-d) shows the corresponding histograms with the distributions of probability. The definition of the probability is $P = N_0/N_t$, where $N_0$ is the number of the initial conditions $(\mathcal{R}, \mathcal{I})$ which display true convergence, while $N_t$ is the number of the total nodes on the plane of complex numbers.

The histograms shown in panels (a-d) of Fig. \ref{hist} can be used for extracting more information, regarding the convergence properties of the NR method. For instance, we can use the Laplace distribution for obtaining the best fits of the right-hand sides (tails) of the histograms (see blue solid lines). We choose to use the Laplace distribution because this is the most natural choice particularly in systems displaying transient chaos (e.g., \cite{ML01,SASL06,SS08}).

The Laplace probability density function (PDF) is given by
\begin{equation}
P(N | l,d) = \frac{1}{2d}
 \begin{cases}
      \exp\left(- \frac{l - N}{d} \right), & \text{if } N < l \\
      \exp\left(- \frac{N - l}{d} \right), & \text{if } N \geq l
 \end{cases},
\label{pdf}
\end{equation}
where the parameters $l$ and $d > 0$ are the location parameter and the parameter of diversity, respectively. From the PDF we need only the $N \geq l$ part because the Laplace distributions refer only to the tails of the probability histograms.

\begin{figure*}[!t]
\centering
\resizebox{\hsize}{!}{\includegraphics{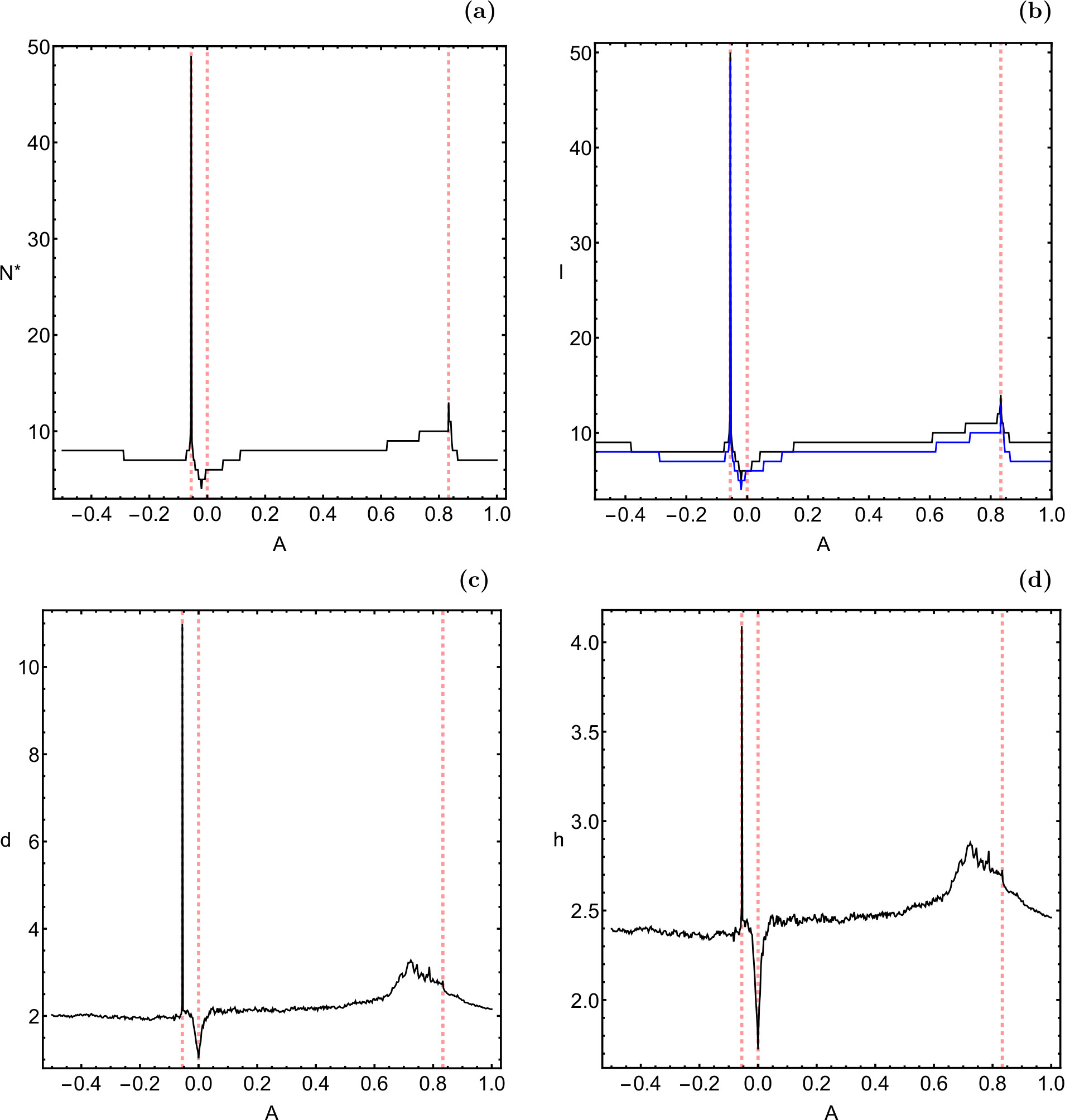}}
\caption{Parametric evolution of the (a-upper left): most probable number of iterations $N^{*}$; (b-upper right): location parameter $l$; (c- lower left): diversity $d$; (d-lower right): differential entropy $h$, as a function of the oblateness $A$. With blue color in panel (b) we denote the parametric evolution of $N^{*}$. The red lines indicate the three critical levels of the oblateness. (Color figure online).}
\label{stats}
\end{figure*}

\begin{figure}[!t]
\centering
\resizebox{0.6\hsize}{!}{\includegraphics{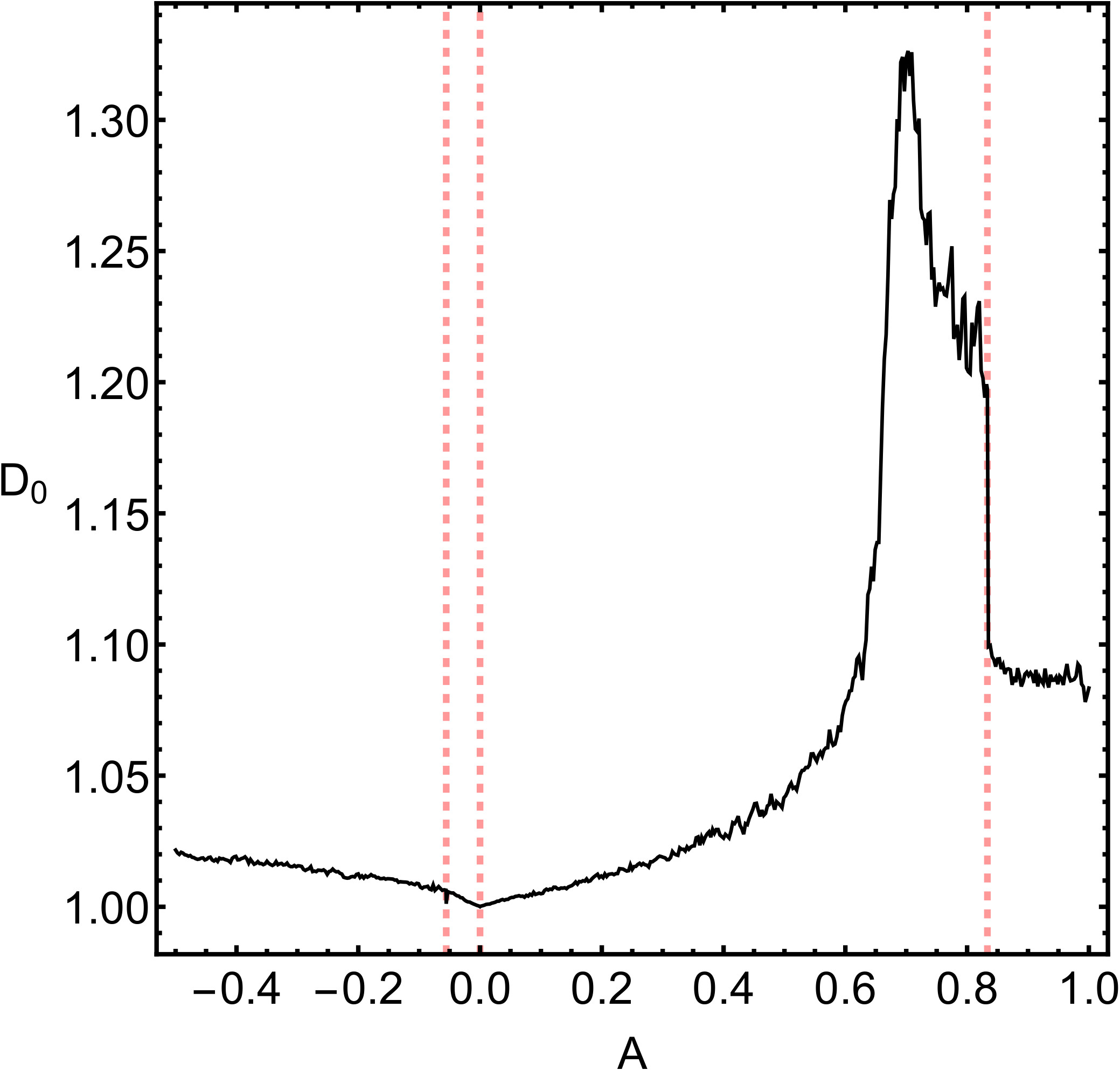}}
\caption{Parametric evolution of the uncertainty or fractal dimension $D_0$, as a function of the oblateness $A$. The red lines indicate the three critical values of $A$. (Color figure online).}
\label{d0}
\end{figure}

\begin{figure*}[!t]
\centering
\resizebox{\hsize}{!}{\includegraphics{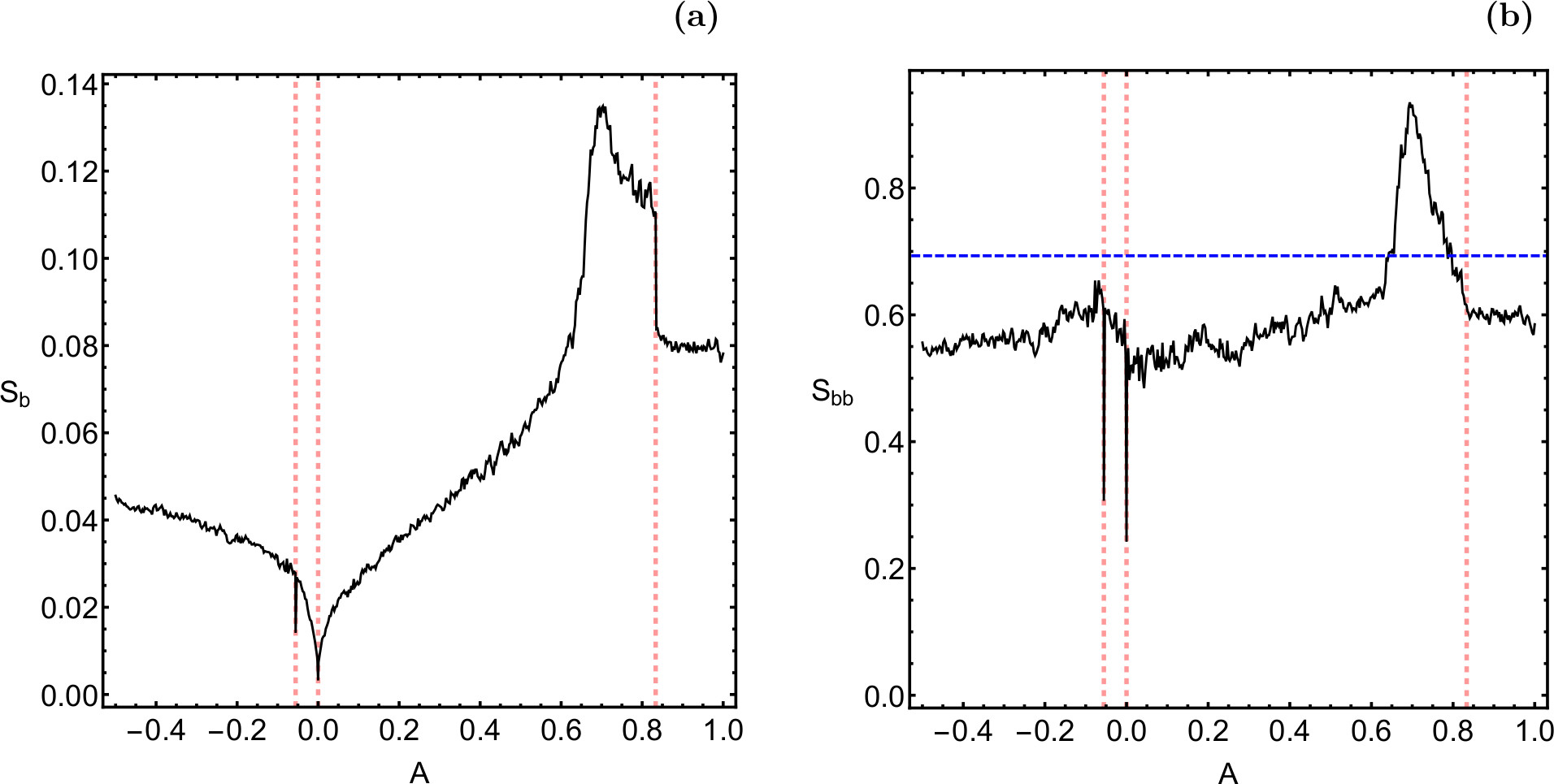}}
\caption{Parametric evolution of the (a-left): entropy of basins $S_b$ and (b-right): boundary entropy of basins $S_{bb}$, as a function of the oblateness $A$. The red lines indicate the three critical values of $A$, while the blue line denotes the critical value $\log 2$. (Color figure online).}
\label{be}
\end{figure*}

\begin{figure*}[!t]
\centering
\resizebox{\hsize}{!}{\includegraphics{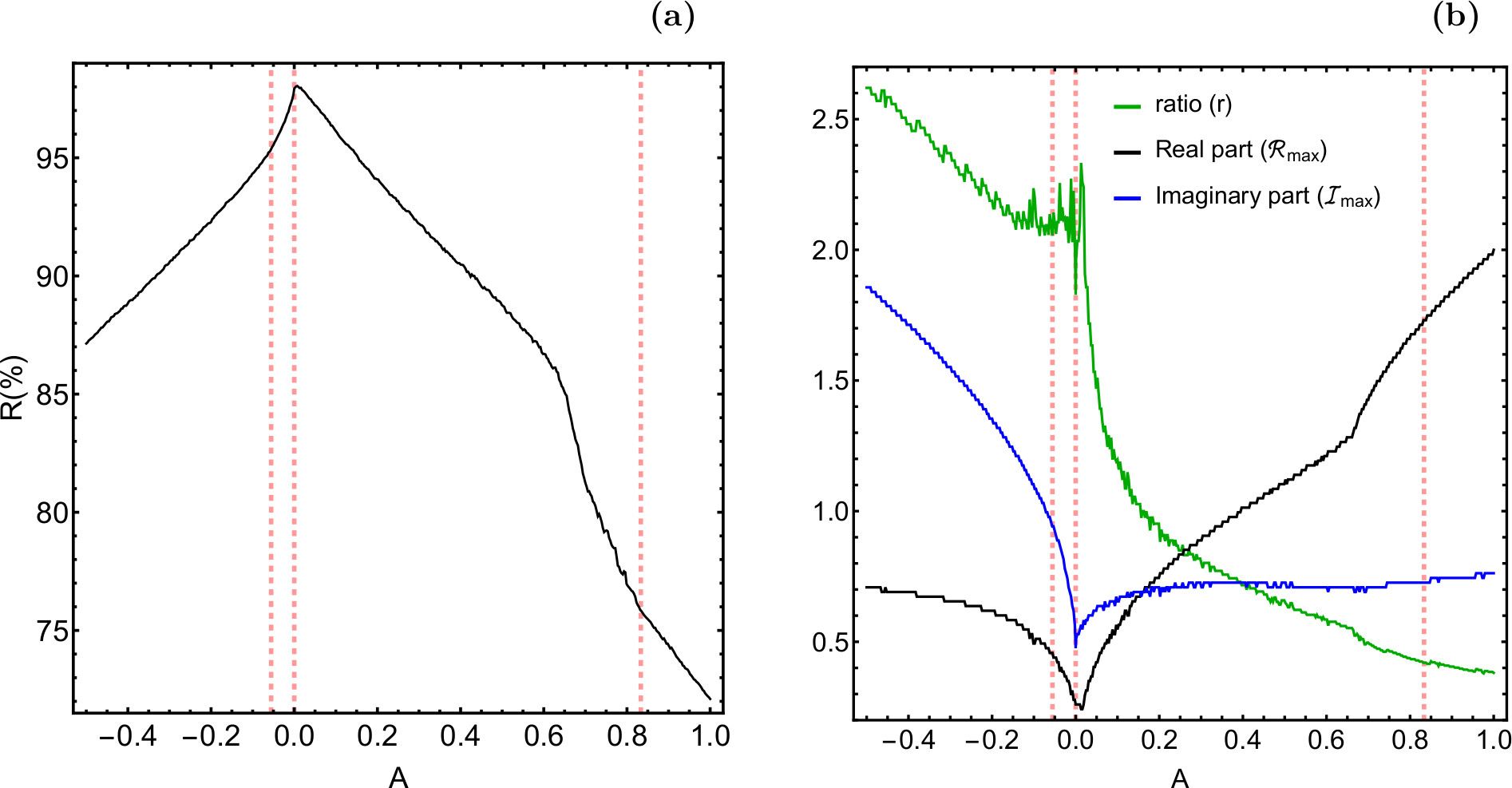}}
\caption{Parametric evolution of the (a-left): area $R$ on the plane of complex numbers, covered by starting points for which the NR method diverges to infinity and (b-right): $\mathcal{R}_{max}$, $\mathcal{I}_{max}$ and $r$, of the overall shape of the Boc. The vertical, dashed, red lines indicate the three critical values of $A$. (Color figure online).}
\label{stats2}
\end{figure*}

In \citetalias{ZSAS18} we investigated using numerical methods the convergence dynamics of the Sitnikov problem with non-spherical primaries however for some individual values of the oblateness $A$. In the present work, we will perform a more systematic numerical analysis in an attempt to determine how $A$ affects the convergence properties of the system. For this task, we classified 1000 grids of $1024 \times 1024$ starting points $(\mathcal{R}, \mathcal{I})$ inside the square region $R = [-2,2] \times [-2,2]$ on the plane of complex numbers, for the range $A \in [-0.5,1]$, thus following the pioneer works of Nagler \cite{N04,N05}. In our calculations the desired accuracy, regarding the coordinates of the attractor, was set to $10^{-15}$, while the maximum allowed number of iterations was $N_{\rm max} = 500$.

The evolution of the most probable number of iterations $N^{*}$, per grid, is illustrated in panel (a) of Fig. \ref{stats}. It is seen, that around the critical levels $A = -1/18$ and $A = 5/6$ we have two peaks, while on the other hand around the critical level $A = 0$ we observe the lowest value of $N^{*}$. In \citetalias{ZSAS18} we had discussed (see bottom row of Fig. 4) that when $A = -1/18$ for a set of starting points the NR scheme requires a significant number of iterations for converging to the attractor $\mathz = 0$. This is exactly why in part (a) of Fig. \ref{stats} we see that the highest value of $N^{*}$ is measured near $A = -1/18$.

In parts (b) and (c) of Fig. \ref{stats} we present the parametric evolution of the location parameter $l$ and the diversity $d$, respectively, as a function of the oblateness $A$. In part (b) we also included, for comparison reasons, using blue color the evolution of the $N^{*}$ of iterations. One can observe that in general terms the location parameter almost coincides with the average number of iterations (almost always $|l - < N >| \leq 2$). This implies that the Laplace probability density function (PDF) can satisfactorily fit the tails of the probability histograms. According to the plot shown in part (c) of Fig. \ref{stats} the diversity is, in most of the cases, low $(d < 3)$, thus indicating the dispersion of the values of $N$ is very close to $N^{*}$. On the other hand, in the vicinity of the critical levels of the oblateness $A = -1/18$ and $A = 0$ the diversity exhibits a local maximum and a local minimum, respectively. Part (d) of Fig. \ref{stats} illustrates the evolution of the differential entropy $h = 1 + \ln(2d)$, where $d$ is the diversity. One can see, that the evolution of both $d$ and $h$ displays similar overall patterns.

The numerical analysis presented in \citetalias{ZSAS18} revealed the fractal regions on the plane of complex numbers. One of the most convenient ways of measuring the degree of fractality of a system is by computing the uncertainty or fractal dimension $D_0$ (see e.g., \cite{O93}), thus following the computational methodology used in \cite{AVS01,AVS09}. Fig. \ref{d0} shows the parametric evolution of the uncertainty dimension, as a function of the oblateness $A$. In the first interval of $\delta$ the fractal dimension is very close to 1, which implies zero fractality. In the first two intervals $D_0$ decreases, while in the third interval its value is mostly reduced. It is seen, that $D_0$ displays the maximum value near the value $A = 0.7$, while the lowest value is measured at $A = 0$.

Another efficient way for quantitatively measuring the degree of fractality of a system is by computing the so-called basin entropy \cite{DWGGS16,DWGGS18}. This method determines the fractality of a basin diagram by the process of examination of its topological properties. The evolution of the entropy of the basins $S_b$, as a function of $A$, is illustrated in panel (a) of Fig. \ref{be}. Once more, we note that in the vicinity of all the critical values of $A$ there are three local minima of $S_b$, mainly because for these values of $A$ the total number of the numerical attractors decreases from 5 to three (when $A = -1/18$ and $A = 5/6$) and one (when $A = 0$). The maximum value of $S_b$ was measured near $A = 0.7$. Therefore, we may argue that two different methods (i.e., the uncertainty dimension and the basin entropy) indicate that the degree of the fractality of the Boc on the plane of complex numbers is maximum near the same value of $A$.

Apart from the basin entropy there is also the boundary basin entropy $S_{bb}$ \cite{DWGGS16}, from which we can extract additional information about the degree of fractality of the Boc. The parametric evolution of $S_{bb}$ is given in part (b) of Fig. \ref{be}. From this type of plot we can also deduce information regarding the fractality of the Boc on the plane of complex numbers. More specifically, we can use the so-called ``log 2 criterion", according to which if $S_{bb} > \log 2$ then the basin boundaries are certainly fractal (here note that the converse statement is not valid). As it is seen in part (b) of Fig. \ref{be} the basin boundaries are certainly fractal only when $0.65 < A < 0.78$. Once more, the lowest values of $S_{bb}$ are reported in the vicinity of the critical values of the oblateness.

At this point, we would like to briefly discuss the efficiency of the NR method. The classification of the 1000 grids of initial conditions suggested that at least for this dynamical system, the numerical method does display an ill behaviour, for some specific sets of initial conditions. In particular, for a large amount of initial conditions the iterator quickly diverges to extremely large complex numbers. In part (a) of Fig. \ref{stats2} we give the parametric evolution of the area $R$ on the plane of complex numbers, covered by starting points which diverge to infinity. It is seen, that the highest value of $R$ is observed at $A = 0$, while for higher values of $A$ the value of the area decreases.

Before closing this section, it would be very illuminating to discuss how the oblateness influences the geometry of the Boc. As we have already seen, the converge regions on the complex plane have finite area. Therefore, we define as $\mathcal{R}_{max}$ and $\mathcal{I}_{max}$ the maximum values of the Boc, along both axes, respectively, while the ratio $r$ is defined as $\mathcal{I}_{max}/\mathcal{I}_{max}$. Part (b) of Fig. \ref{stats2} shows the evolution of $\mathcal{R}_{max}$, $\mathcal{I}_{max}$ and $r$, as a function of the oblateness. One can see, that for negative values of $A$ the overall structure of the Boc is elongated along the vertical axis. On the contrary, for $A > 0$ the value of $\mathcal{R}_{max}$ increases and at about $A = 0.15$ we have that $\mathcal{R}_{max} =  \mathcal{I}_{max}$. Moreover, for $A > 0.15$ the value of $\mathcal{R}_{max}$ increases rapidly which implies that the overall shape of the convergence regions becomes elongated along the horizontal axis.

\section{Concluding remarks}
\label{conc}

The present paper can be considered as a continuation of \citetalias{ZSAS18}. The scope of this work was to study the convergence dynamics of the Sitnikov problem with non-spherical primaries. Using the NR iterative method we revealed the Boc on the plane of complex numbers, by means of color-coded diagrams. Moreover, we demonstrated how the oblateness $A$ influences the speed and the accuracy of the method. At the same time, it was determined how the same parameter affects the degree of fractality of the convergence regions, by computing modern quantitative indices, such as the uncertainty dimension and the (boundary) basin entropy.

In this work we demonstrate for the first time how the oblateness of the Sitnikov problem with spheroidal primaries influences the overall properties of the system. Additionally, we relate also for the first time different techniques for measuring the fractal degree of a dynamical system. More specifically, we computed and compared the results of both the fractal dimension and the (boundary) basin entropy. On this basis, we claim that the resented outcomes of the article are interesting and novel and add new information on the field of the convergence properties of Hamiltonian systems.

The most important findings of our numerical exploration are listed here:
\begin{enumerate}
  \item For the majority of the studied cases, regarding the value of $A$, the NR iterative scheme requires an average number of 8 iterations for leading to one of the numerical attractors (roots), while only near the critical values of $A$ the average number of iterations increases.
  \item Our numerical analysis reported the existence of only starting points for which the numerical iterator quickly diverges to infinity. On the other hand, there is no indication of false or non-converging starting points on the plane of complex numbers.
  \item The highest fractal degree (measured by using the uncertainty dimension as well as the entropy of the basins) of the convergence diagrams on the plane of complex numbers correspond to about $A = 0.7$, while the lowest values of the degree of fractality were measured near the critical levels of $A$, where the number of the numerical attractors is reduced.
  \item Exploiting the information of the computation of the boundary basin entropy and the ``log 2 criterion" we proved that in this system the basin boundaries of the convergence regions on the plane of complex numbers are certainly fractal for $0.65 < A < 0.78$.
  \item For negative values of the oblateness the overall shape of the convergence regions on the plane of complex numbers is elongated along the vertical axis, while for $A > 0.15$ the geometry changes and the shape becomes elongated along the horizontal axis.
  \end{enumerate}

The numerical routine of the NR iterative method was written in the standard version of \verb!FORTRAN 77! \cite{PTVF92}. For classifying the nodes on the plane of complex numbers, roughly about 2.5 minutes of CPU time, per grid, was required using an Quad-Core Intel i7 vPro 4.0 GHz processor. The version 11.3 of Mathematica$^{\circledR}$ \cite{W03} has been deployed for constructing all the graphics of the paper.



\footnotesize
\section*{Compliance with Ethical Standards}

\begin{itemize}
  \item Funding: The authors state that they have not received any research grants.
  \item Conflict of interest: The authors declare that they have no conflict of interest.
\end{itemize}

\end{document}